\documentclass[conference]{IEEEtran}

\usepackage{amsmath, amsfonts} 
\usepackage{amsthm}
\usepackage[cmintegrals]{newtxmath}
\usepackage{graphicx}
\usepackage{multirow}

\newcommand{\overbar}[1]{\mkern 1.5mu\overline{\mkern-1.5mu#1\mkern-1.5mu}\mkern 1.5mu}

\newtheorem{newdef}{Definition}
\newtheorem{theorem}{Theorem}

\begin{document}

\title{Spark Optimization of Linear Codes for Reliable Data Delivery by Relay Drones}

\author{\IEEEauthorblockN{Ioannis~Chatzigeorgiou}
\IEEEauthorblockA{School of Computing and Communications\\
Lancaster University, United Kingdom\\
Email: i.chatzigeorgiou@lancaster.ac.uk}}

\maketitle
\begin{abstract}
Data gathering operations in remote locations often rely on relay drones, which collect, store and deliver transmitted information to a ground control station. The probability of the ground control station successfully reconstructing the gathered data can be increased if random linear coding (RLC) is used, especially when feedback channels between the drones and the transmitter are not available. RLC decoding can be complemented by partial packet recovery (PPR), which utilizes sparse recovery principles to repair erroneously received data before RLC decoding takes place. We explain that the spark of the transpose of the parity-check matrix of the linear code, that is, the smallest number of linearly-dependent columns of the matrix, influences the effectiveness of PPR. We formulate a spark optimization problem and obtain code designs that achieve a gain over PPR-assisted RLC, in terms of the probability that the ground control station will decode the delivered data.\vspace{2pt}
\end{abstract}



\section{Introduction}
\label{sec.intro}

Unmanned aerial vehicles, commonly referred to as \textit{drones}, are used in surveillance, reconnaissance, rescue and other military operations. Relay drones, in particular, have the potential to lower the demand for limited satellite resources, as they can provide on-demand wireless connectivity between remote network nodes when the line-of-sight channel is blocked, e.g. due to physical obstacles, or is compromised, e.g. due to intentional jamming or unintentional interference~\cite{Carr2009}. In delay-tolerant applications, relay drones are not always required to establish an end-to-end communication path. They could be dispatched to a remote site with the objective of collecting information transmitted by an on-site data-gathering node. For example, the node could be a reconnaissance team sending critical data for a military operation. Collected information would be stored on the drones, which would carry and safely deliver it to a ground control station for further processing.

This work considers a transmitter broadcasting information to drones in the absence of feedback. In~\cite{Chatzigeorgiou2020}, we quantified the performance gain of \textit{random linear coding} (RLC)~\mbox{\cite{Byers1998,Ho2006}} over the sequential transmission and periodic repetition of data packets, known as a \textit{data carousel}, in terms of the probability that a ground control station will recover the transmitted information. RLC is an application-layer forward error correction scheme (AL-FEC), which generates coded packets that are random linear combinations of data packets. RLC decoding discards received coded packets that have been corrupted by errors and attempts to reconstruct the data packets from correctly received coded packets. Mohammadi \textit{et~al.}~\cite{Mohammadi2016} proposed partial packet recovery (PPR), which utilizes corrupted coded packets, complements RLC decoding and has the potential to improve the chances of data packet recovery. The motivation for this paper is to revisit PPR, identify factors that influence the effectiveness of PPR, and develop PPR-assisted AL-FEC schemes for challenging channel conditions that are often observed in military communications.

The remainder of this paper has been organized as follows: Section~\ref{sec.System} presents the system model, introduces the notation, briefly describes RLC encoding and decoding, and explains how PPR can improve the reliability of the system. Section~\ref{sec.Syndrome_Decoding} delves into PPR-assisted RLC decoding and discusses two common approaches for repairing corrupted packets that would have otherwise been discarded by the RLC decoder. Section~\ref{sec.Design} proposes code design rules that can improve the ability of PPR to repair packets. Performance comparisons are presented in Section~\ref{sec.Results} and key findings are summarized in Section~\ref{sec.Conclusions}.


\begin{figure}[t]
\centering
\includegraphics[width=0.84\columnwidth]{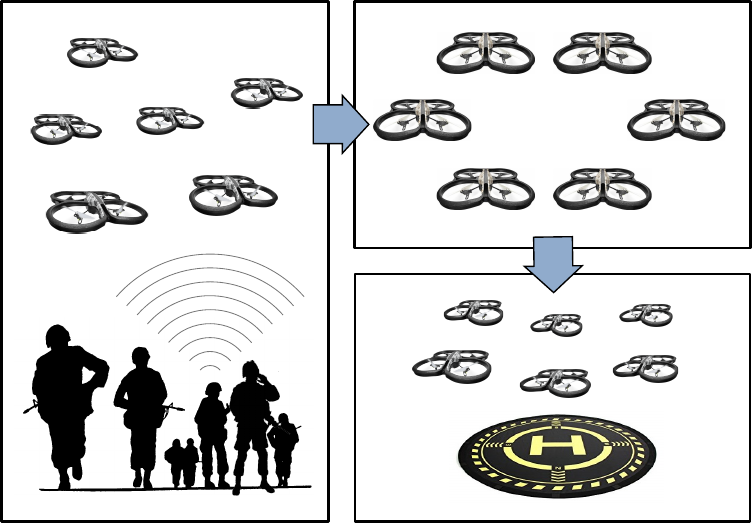}
\caption{Sensitive information is encoded using random linear coding and is broadcast to $M=6$ drones. The drones collect, store and deliver the coded information to the ground control station for decoding.}
\label{fig:system_model}
\end{figure}

\section{System Model and Problem Formulation}
\label{sec.System}

We consider $M$ drones traveling from a ground control station to a site, where a transmitter broadcasts sensitive information to the $M$ drones, as shown in Fig.~\ref{fig:system_model}. Before broadcasting, the information is segmented into $K$ source packets, $\mathbf{u}_1,\ldots,\mathbf{u}_K$. A source packet $\mathbf{u}_k$, for $k=1,\dots,K$, is modeled as a sequence of $L$ symbols from a finite field of size $q$, that is, $\mathbf{u}_k\in\mathbb{F}^{L}_{q}$, where $q$ is a prime power. The $K$ source packets are encoded into $N\geq K$ coded packets using RLC~\cite{Ho2006}. At time step $n$, the transmitter generates coded packet $\mathbf{x}_n\in\mathbb{F}^{L}_{q}$, which is a random linear combination of the $K$ source packets. The coded packet $\mathbf{x}_{n}$ can thus be expressed as:
\begin{equation}
\label{eq.RLNC_packet}
\mathbf{x}_{n} = \sum_{k=1}^{K} g_{n,k}\,\mathbf{u}_k,\;\;\mathrm{for}\;\; n=1,\dots,N,
\end{equation}
where each coefficient $g_{n,k}$ is chosen uniformly at random from $\mathbb{F}_{q}$. Using matrix notation, expression $\eqref{eq.RLNC_packet}$ can be rewritten as:
\begin{equation}
\label{eq.RLNC_matrix}
\mathbf{X} = \mathbf{G}\,\mathbf{U},
\end{equation}
where the matrices $\mathbf{X}\in\mathbb{F}^{N\times L}_{q}$, $\mathbf{G}\in\mathbb{F}^{N\times K}_{q}$ and $\mathbf{U}\in\mathbb{F}^{K\times L}_{q}$ have been obtained as follows:
\begin{equation}
\label{eq.Matrix_definitions}
\mathbf{X}=\left[\!\begin{array}{c} \mathbf{x}_{1}\\ \vdots\\ \mathbf{x}_{N} \end{array}\!\!\right],\,%
\mathbf{G}=\left[\! \begin{array}{ccc}
g_{1,1} & \!\cdots & \!g_{1,K} \\
\vdots  & \!\ddots & \!\vdots  \\
g_{N,1} & \!\cdots & \!g_{N,K} \end{array} \!\!\right]\,%
\textrm{and}\,\;%
\mathbf{U}=\left[\!\begin{array}{c} \mathbf{u}_{1}\\ \vdots\\ \mathbf{u}_{K} \end{array}\!\!\right].
\end{equation}
Matrix $\mathbf{G}$ is referred to as the \textit{generator matrix}. In practice, as explained in \cite{Mohammadi2016}, a seed is used to initialise the pseudo-random number generator that outputs the values of the elements in $\mathbf{G}$. Given that the value of the seed can be conveyed in the packet headers, we assume that both the transmitter and the ground control station have knowledge of $\mathbf{G}$.

A Cyclic Redundancy Check (CRC) is calculated and appended to each coded packet. The $N$ coded packets are sent to the $M$ drones over broadcast erasure channels. The link between the transmitter and drone $\mathrm{D}_{m}$, for $m=1,\ldots,M$, is characterized by packet erasure probability $\varepsilon_m$. Each link encompasses the wireless channel but also the physical layer of the transmitter and the receiving drone. The packet erasure probability $\varepsilon_m$ captures the channel conditions as well as the modulation and coding scheme (MCS) employed by the transmitter and drone $\mathrm{D}_{m}$. It can often be expressed in analytic form as a function of the channel parameters and a Signal-to-Noise Ratio (SNR) threshold that is specific to the adopted MCS. For instance, if the Nakagami fading model accurately describes the channel between the transmitter and drone $\mathrm{D}_{m}$, as in \cite{Yanmaz2013}, the value of $\varepsilon_m$ can be computed using \cite[eq.~(12)]{Xi2011}.

We assume that feedback channels between the drones and the transmitter are not available. This could be the case when the communication protocol does not support feedback loops, or when the feedback links of the power-constrained drones are rendered unreliable by intentional jamming.

When transmission has been completed, the drones do not attempt to reconstruct the source information but return to the ground control station in order to deliver the collected coded packets. Received copies of coded packet $\mathbf{x}_n$ may be available on one or more drones. The copy that will be stored on the ground control station, denoted by $\mathbf{y}_n$, should be free of errors and, hence, satisfy the CRC; in this case $\mathbf{y}_n=\mathbf{x}_n$. Otherwise, if none of the drones carry a copy of $\mathbf{x}_n$ that satisfies the CRC, one of the erroneous copies will be randomly chosen and deposited on the ground control station; when the stored packet has been corrupted by errors, we can write $\mathbf{y}_n=\mathbf{x}_n+\mathbf{e}_n$. In the latter case, the row vector $\mathbf{e}_n\in\mathbb{F}^L_q$ contains non-zero elements in the positions where errors have occurred and zero elements in the remaining positions.

Let $R$ be the ordered set of indices $n$ for which $\mathbf{y}_n=\mathbf{x}_n$ and $\mathbf{e}_n=\mathbf{0}$, where $R\subseteq\{1,\ldots,N\}$ and $\vert R\vert=N_R\!\leq\! N$. The received coded packets $\mathbf{y}_1,\ldots,\mathbf{y}_N$ form the rows of the $N\times L$ matrix $\mathbf{Y}$, while the row vectors $\mathbf{e}_1,\ldots,\mathbf{e}_N$ compose the $N\times L$ error matrix $\mathbf{E}$. The correctly received coded packets $\mathbf{y}_i$, for all $i\in R$, make the rows of the $N_R\times L$ matrix $\mathbf{Y}_R$. On the other hand, the erroneously received coded packets $\mathbf{y}_j$ have indices in the set $\overbar{R}=\{1,\ldots,N\}\backslash R$ and make the rows of the $(N-N_R)\times L$ matrix $\mathbf{Y}_{\overbar{R}}$. In a similar manner, the sets $R$ and $\overbar{R}$ can be used to obtain $\mathbf{X}_R$, $\mathbf{X}_{\overbar{R}}$, $\mathbf{G}_R$, $\mathbf{G}_{\overbar{R}}$ and $\mathbf{E}_{\overbar{R}}$. The relationships between these matrices are summarized below:
\begin{equation}
\label{eq.Matrices_at_GCS}
\mathbf{Y} = \mathbf{X}+\mathbf{E}\Leftrightarrow \left\{
\begin{array}{l}
      \mathbf{Y}_R=\mathbf{X}_R=\mathbf{G}_R\mathbf{U},\\
      \mathbf{Y}_{\overbar{R}}=\mathbf{X}_{\overbar{R}}+\mathbf{E}_{\overbar{R}}.\\
\end{array}
\right.
\end{equation}

RLC decoders utilize the $N_R$ correctly received coded packets, represented by $\mathbf{Y}_R$. The $N-N_R$ erroneous coded packets are assumed to be `erased' by the channel and $\mathbf{Y}_{\overbar{R}}$ is discarded. When $N_R\geq K$, the RLC decoder will check if $K$ of the $N_R$ rows of $\mathbf{G}_R$ are linearly independent or, equivalently, if the rank of $\mathbf{G}_R$ is $K$. If $\mathrm{rank}(\mathbf{G}_R)=K$, then $\mathbf{Y}_R=\mathbf{G}_R\mathbf{U}$ can be seen as a linear system of $N_R\geq K$ equations, which can be reduced to a system of $K$ linearly independent equations with $K$ unknowns (i.e., source packets). This $K\times K$ system can be solved, e.g., using Gaussian elimination, and return a unique solution for $\mathbf{U}$. Otherwise, if $\mathrm{rank}(\mathbf{G}_R)<K$, the linear system is underdetermined, i.e., it has fewer linearly independent equations than unknowns, therefore a unique solution cannot be obtained.

As Mohammadi~\textit{et al.} demonstrated in \cite{Mohammadi2016}, the $N-N_R$ partially corrupted coded packets often contain large error-free segments, even in challenging channel conditions. Partial packet recovery (PPR) methods exploit the sparse distribution of errors and the fact that the generator matrix $\mathbf{G}$ is known at the ground control station, and attempt to repair received coded packets that contain errors. Those rows of $\mathbf{Y}_{\overbar{R}}$ that correspond to repaired coded packets are updated and moved to $\mathbf{Y}_R$ in an effort to increase the chances of the RLC decoder recovering the $K$ source packets. We adapt the PPR process presented in \cite{Mohammadi2016} to the system model under consideration and describe it in more detail in the following section.


\section{PPR-assisted RLC Decoding}
\label{sec.Syndrome_Decoding}

The authors in \cite{Mohammadi2016} employed \textit{systematic} RLC, according to which the first $K$ of the $N$ packets that are pushed down to the physical layer of the transmitter are identical to the $K$ source packets, and only the remaining $N-K$ packets are random linear combinations of the source packets. As a result, the generator matrix $\mathbf{G}$ can be expressed in standard form:
\begin{equation}
\label{eq.sys_generator_matrix}
\mathbf{G}=
\left[\!\begin{array}{c} \mathbf{I}_K\\ \mathbf{P}\end{array}\!\right],
\end{equation}
where $\mathbf{I}_K$ is the $K\times K$ identity matrix and $\mathbf{P}$ is a $(N-K)\times K$ randomly generated matrix with elements from $\mathbb{F}_q$.

Knowledge of the $N\times K$ generator matrix $\mathbf{G}$ enables the ground control station to derive the $N\times (N-K)$ parity-check matrix $\mathbf{H}$, as follows:
\begin{equation}
\label{eq.sys_parity_check_matrix}
\mathbf{H}=\left[\!\begin{array}{c} -\mathbf{P}\;\vert\;\mathbf{I}_{N-K}\end{array}\!\right]^\top,
\end{equation}
so that:
\begin{equation}
\label{eq.zero_product}
\mathbf{H}^\top\,\mathbf{G}=\mathbf{0}.
\end{equation}
Matrix negation in \eqref{eq.sys_parity_check_matrix} is performed in $\mathbb{F}_q$, e.g., $-\mathbf{P}=\mathbf{P}$ in $\mathbb{F}_2$. Multiplication of $\mathbf{H}^\top$ by the received matrix $\mathbf{Y}$ produces the $(N-K)\times L$ \textit{syndrome matrix} $\mathbf{S}$, i.e., $\mathbf{S} = \mathbf{H}^\top \mathbf{Y}$. Using \eqref{eq.zero_product}, we find that the relationship between the syndrome matrix $\mathbf{S}$ and the error matrix $\mathbf{E}$ is:
\begin{equation}
\label{eq.syndrome_decoding_full}
\begin{split}
\mathbf{S} & = \mathbf{H}^\top \mathbf{Y} \\
 & = \mathbf{H}^\top (\mathbf{G}\mathbf{U}+\mathbf{E})\\
 & = \mathbf{H}^\top \mathbf{E}.
\end{split}
\end{equation}
We mentioned in Section~\ref{sec.System} that the set $R$ contains the $N_R$ indices of the correctly received packets, which correspond to all-zero rows in matrix $\mathbf{E}$. All-zero rows in $\mathbf{E}$ do not contribute to the generation of $\mathbf{S}$. Therefore, the $N_R$ all-zero rows of $\mathbf{E}$ with indices in $R$ can be removed in order to reduce the $N\times L$ matrix $\mathbf{E}$ to the $(N-N_R)\times L$ matrix $\mathbf{E}_{\overbar{R}}$. In a similar manner, $\mathbf{H}$ reduces to $\mathbf{H}_{\overbar{R}}$ and expression \eqref{eq.syndrome_decoding_full} assumes the form:
\begin{equation}
\label{eq.syndrome_decoding_partial}
\mathbf{S} = \left(\mathbf{H}_{\overbar{R}}\right)^{\!\top} \mathbf{E}_{\overbar{R}}.
\end{equation}
If we denote by $[\mathbf{S}]_{*,j}$ and $[\mathbf{E}_{\overbar{R}}]_{*,j}$ the $j$-th column of matrices $\mathbf{S}$ and $\mathbf{E}_{\overbar{R}}$, respectively, we can rewrite \eqref{eq.syndrome_decoding_partial} as:
\begin{equation}
\label{eq.syndrome_decoding_partial_per_col}
\left[\mathbf{S}\right]_{*,j} = \left(\mathbf{H}_{\overbar{R}}\right)^{\!\top} \left[\mathbf{E}_{\overbar{R}}\right]_{*,j}\quad\text{for}\quad j=1,\ldots,L.
\end{equation}
Expression \eqref{eq.syndrome_decoding_partial_per_col} represents $L$ independent systems of $N-K$ linear equations with $N-N_R$ unknowns per equation.

The ground control station uses the $N$ stored coded packets to construct matrices $\mathbf{Y}_R$ and $\mathbf{Y}_{\overbar{R}}$, as shown in \eqref{eq.Matrices_at_GCS}. If matrix $\mathbf{G}_R$ has rank $K$, the $K$ source packets can be obtained from $\mathbf{Y}_R$, as explained in Section~\ref{sec.System}. Otherwise, the ground control station computes $\mathbf{S}$ and $\mathbf{H}_{\overbar{R}}$, and attempts to estimate all columns of $\mathbf{E}_{\overbar{R}}$ so that equation \eqref{eq.syndrome_decoding_partial_per_col} is satisfied. If $N_R<K$, the system of linear equations in \eqref{eq.syndrome_decoding_partial_per_col} is underdetermined. If we take into account that partially corrupted packets often contain large error-free segments, we can infer that $\mathbf{E}_{\overbar{R}}$ is a \textit{sparse} matrix, that is, $\mathbf{E}_{\overbar{R}}$ contains more zero elements than non-zero elements. Based on this observation, the solution to \eqref{eq.syndrome_decoding_partial_per_col} can be formulated as an `$\ell_0$ minimization' problem:
\begin{equation}
\label{eq.L0_minimization}
\begin{split}
\left[\hat{\mathbf{E}}_{\overbar{R}}\right]_{*,j} =  &\arg\min_{\mathbf{w}^{\top}}\;\lVert \mathbf{w} \rVert_0 \\
&\text{subject to}\; \left(\mathbf{H}_{\overbar{R}}\right)^{\!\top}\mathbf{w}^{\top}=\left[\mathbf{S}\right]_{*,j}
\end{split}
\end{equation}
where $\mathbf{w}\in\mathbb{F}^{N-N_R}_q$ is a row vector that should satisfy the constraint in \eqref{eq.L0_minimization} and have the minimum possible number of non-zero elements. To count the non-zero elements in $\mathbf{w}$, the $\ell_0$ norm is used, which is defined as $\lVert \mathbf{w} \rVert_0=\lvert w_1\rvert^0+\ldots+\lvert w_{N-N_R}\rvert^0$ assuming that $0^0=0$ \cite{Donoho2001}.

In \cite{Mohammadi2016}, the authors propose two approaches for the evaluation of  $\mathbf{\hat{E}}_{\overbar{R}}$. The first approach, which is inspired by the compressed sensing (CS) literature, replaces the non-convex $\ell_0$ norm with the convex $\ell_1$ norm, solves the optimization problem over the set of real numbers, and rounds off the derived values to the nearest elements of $\mathbb{F}_q$. The second approach, referred to as \textit{syndrome decoding}, considers \eqref{eq.L0_minimization} and initiates an exhaustive search for a candidate solution; the sparsity of the row vector $\mathbf{w}$ is gradually reduced, i.e., the number of non-zero elements in $\mathbf{w}$ increases, and the search concludes when the sparsest vector $\mathbf{w}$ that satisfies the constraint in \eqref{eq.L0_minimization} has been identified.

When candidate solutions for the $L$ columns of $\hat{\mathbf{E}}_{\overbar{R}}$ have been obtained and $\hat{\mathbf{E}}_{\overbar{R}}$ has been constructed, an estimate of $\mathbf{X}_{\overbar{R}}$, denoted by $\hat{\mathbf{X}}_{\overbar{R}}$, can be derived using \eqref{eq.Matrices_at_GCS}:
\begin{equation}
\label{eq.XR_estimate}
\hat{\mathbf{X}}_{\overbar{R}}=\mathbf{Y}_{\overbar{R}}-\hat{\mathbf{E}}_{\overbar{R}}.
\end{equation}
To determine which of the estimated rows of $\hat{\mathbf{X}}_{\overbar{R}}$ correspond to successfully repaired packets, CRC verification for each row is carried out. Let $\nu$ denote the number of rows in $\hat{\mathbf{X}}_{\overbar{R}}$ that passed CRC verification. The indices of the $\nu$ repaired packets are removed from set $\overbar{R}$ and added to set $R$, and the corresponding rows of $\hat{\mathbf{X}}_{\overbar{R}}$ are moved to $\mathbf{Y}_R$. The cardinalities of sets $\overbar{R}$ and $R$ change to $N-N_R-\nu$ and $N_R+\nu$, respectively, while the dimensions of $\hat{\mathbf{X}}_{\overbar{R}}$ and $\mathbf{Y}_R$ change to $(N-N_R-\nu)\times L$ and $(N_R+\nu)\times L$, respectively. The indices of the $\nu$ repaired packets are also used to identify the rows of the generator matrix $\mathbf{G}$ that should be appended to $\mathbf{G}_R$. If the $\nu$ repaired packets increase the rank of the enlarged $(N_R+\nu)\times K$ matrix $\mathbf{G}_R$ to $K$, PPR will have been successful in assisting the RLC decoder to recover the $K$ source packets.

The authors in \cite{Mohammadi2016} consider operations in $\mathbb{F}_2$ and compare the CS-based approach with syndrome decoding in terms of decoding probability and computational complexity. Results demonstrate that when the number of transmitted packets is fixed, the CS-based approach can recover the $K$ source packets with a slightly higher probability than syndrome decoding for a broad range of channel conditions but at a significantly higher computational cost. Given that the timely recovery of sensitive data is of essence in the scenario considered in this paper, we use syndrome decoding in Section~\ref{sec.Results}. Nevertheless, the code design rules that we propose in the following section can improve the ability of PPR to repair corrupted packets, irrespective of the decoding approach.


\section{Code Designs with Optimized Spark}
\label{sec.Design}

The structure of the generator matrix $\mathbf{G}$, shown in \eqref{eq.sys_generator_matrix}, ensures that $\mathrm{rank}(\mathbf{G})=K$. If $\mathbf{G}$ did not have full rank, RLC decoding would have failed to recover the $K$ source packets, even in perfect channel conditions. Whereas the rank of $\mathbf{G}$ is important in RLC decoding, the \textit{spark} of $\mathbf{H}^{\top}$ plays a key role in PPR, as will become clear in this section.

The term `spark' of a matrix was first coined by Donoho and Elad in \cite{Donoho2003} and is defined as follows:
\begin{newdef}[\cite{Elad2010}, p.23]
The spark of a matrix $\mathbf{A}$ is the smallest number of columns from $\mathbf{A}$ that are linearly dependent. Using mathematical notation, we can write
\begin{equation*}
\mathrm{spark}(\mathbf{A}):=\min\{\lVert\mathbf{z}\rVert_0\}\;\,\text{subject to}\;\,\mathbf{A}\mathbf{z}^{\top}=\mathbf{0}\;\,\text{for}\;\,\mathbf{z}\neq\mathbf{0}.
\end{equation*}
\end{newdef}

For example, if matrix $\mathbf{A}\in\mathbb{F}^{\,4\times4}_2$ assumes the form
\begin{equation*}
\label{eq.example_matrix}
\mathbf{A}=\left[\! \begin{array}{cccc}
1 & 1 & 1 & 0 \\
1 & 0 & 1 & 1 \\
0 & 0 & 1 & 0 \\
0 & 0 & 0 & 0
\end{array}\right],
\end{equation*}
the smallest set of linearly dependent columns from $\mathbf{A}$ consists of the 1st, 2nd and 4th columns of $\mathbf{A}$. Modulo-$2$ addition of these three columns gives $\mathbf{0}$, which proves that they are linearly dependent, therefore $\mathrm{spark}(\mathbf{A})=3$. Equivalently, $\mathbf{z}=[1\, 1\, 0\, 1\,]$ is the row vector with the fewest nonzero elements satisfying $\mathbf{A}\mathbf{z}^{\top}=\mathbf{0}$, hence $\mathrm{spark}(\mathbf{A})=\lVert\mathbf{z}\rVert_0=3$. Essentially, the nonzero elements of $\mathbf{z}$ identify the linearly-dependent columns of $\mathbf{A}$.

If we consider a system of linear equations $\mathbf{A}\mathbf{z}^{\top}=\mathbf{b}^{\top}$, the spark of $\mathbf{A}$ gives the following simple criterion for the uniqueness of sparse solutions:
\begin{theorem}[\cite{Elad2010}, p.24]
\label{th.uniqueness}
If a system of equations $\mathbf{A}\mathbf{z}^{\top}=\mathbf{b}^{\top}$ has a solution $\mathbf{z}$ obeying $\lVert \mathbf{z}\rVert_0<\mathrm{spark}(\mathbf{A})/2$, this solution is necessarily the sparsest possible.
\end{theorem}
If our objective is to estimate $\mathbf{z}$, given that $\mathbf{z}$ is a `sufficiently sparse' solution, Theorem~\ref{th.uniqueness} stipulates that the sparsest solution will be correctly identified if $\mathrm{spark}(\mathbf{A})>2\lVert\mathbf{z}\rVert_0$. In our case, we are called to solve $L$ systems of linear equations of the form $\mathbf{H}^\top \mathbf{E}=\mathbf{S}$, as shown in \eqref{eq.syndrome_decoding_full}, where $\mathbf{H}^{\top}$ plays the role of matrix $\mathbf{A}$ in $\mathbf{A}\mathbf{z}^{\top}=\mathbf{b}^{\top}$. The error matrix $\mathbf{E}$ can be accurately estimated, assuming that it is sufficiently sparse, if the spark of $\mathbf{H}^{\top}$ is more than twice the number of nonzero elements in \textit{each} column of $\mathbf{E}$, according to Theorem~\ref{th.uniqueness}. Therefore, maximization of the spark of $\mathbf{H}^{\top}$ will increase the chances of PPR being successful. Given the structure of $\mathbf{H}^{\top}$, presented in \eqref{eq.sys_parity_check_matrix}, the optimization problem takes the form:
\begin{equation}
\label{eq.spark_maximization}
\max_{\tilde{\mathbf{P}}\in\mathbb{F}^{\epsilon\times K}_q}\!\left\{\mathrm{spark}\Big(\big[-\tilde{\mathbf{P}}\;\vert\;\mathbf{I}_{\epsilon}\big]\Big)\right\},
\end{equation}
for a given value of $N=K+\epsilon$. We denote by $Q^{(\epsilon)}$ the set that contains all instances of $\tilde{\mathbf{P}}$ that maximize the spark of $\mathbf{H}^{\top}$. Spark optimization should be performed for all possible values of $N$, i.e., $N=K+1, K+2,\ldots$, and sets $Q^{(1)}, Q^{(2)}, ...$ should be created; when the transmitter assigns a value to $N$, e.g., $N=K+\epsilon$, in the absence of feedback from drones, matrix $\mathbf{P}$ will be drawn from set $Q^{(\epsilon)}$. This is in contrast to RLC, where $\mathbf{P}$ is randomly generated for any value of $N>K$. Recall that $\mathbf{P}$ is used in the construction of $\mathbf{G}$ and $\mathbf{H}^{\top}$ as per \eqref{eq.sys_generator_matrix} and \eqref{eq.sys_parity_check_matrix}.

A key feature of systematic RLC is that the probability of an entry in matrix $\mathbf{P}$ taking a value $\delta\in\mathbb{F}_q$ is uniform and equal to $1/q$. Deviations from this rule have an impact on the decoding probability, the computational complexity and the delay of RLC. For example, if the zero element is favored more than the remaining $q-1$ elements of $\mathbb{F}_q$ when the entries of $\mathbf{P}$ are being set, the decoding complexity of RLC will be decreased but the decoding probability will be reduced \cite{Tassi2015}. To keep the decoding complexity of the proposed high-spark code designs comparable to the decoding complexity of RLC, we give preference to matrices in sets $Q^{(1)}, Q^{(2)},...$, that contain each element of $\mathbb{F}_q$ in similar or equal amounts. To this end, we denote by $\mathrm{PoE}\!\left(\tilde{\mathbf{P}}, \delta\right)$ the function that returns the proportion of entries in $\tilde{\mathbf{P}}$ from $Q^{(\epsilon)}$ that are equal to $\delta\in\mathbb{F}_q$ when $N=K+\epsilon$, and define it as:
\begin{equation}
\label{eq.proportion_matrix}
\mathrm{PoE}\!\left(\tilde{\mathbf{P}},\delta\right):=\frac{1}{\epsilon K}\sum_{i=1}^{\epsilon}\sum_{j=1}^{K} \mathbf{1}_{\delta}\!\left([\tilde{\mathbf{P}}]_{i,j}\right),
\end{equation}
where the indicator function $\mathbf{1}_{\delta}(z)$ returns $\mathbf{1}_{\delta}(z)=1$ if $z=\delta$, otherwise $\mathbf{1}_{\delta}(z)=0$. Based on this definition, we can compute the proportion of $\delta\in\mathbb{F}_q$ in set $Q^{(\epsilon)}$ as follows:
\begin{equation}
\label{eq.proportion_set}
\mathrm{PoE}\!\left(Q^{(\epsilon)},\delta\right):=\frac{1}{|Q^{(\epsilon)}|}\sum_{\tilde{\mathbf{P}}\in Q^{(\epsilon)}}\mathrm{PoE}\!\left(\tilde{\mathbf{P}},\delta\right),
\end{equation}
where the sum is taken over all matrices in $Q^{(\epsilon)}$.

If our objective is to maximize the ability of PPR to repair corrupted packets without deviating greatly from the decoding complexity of RLC\footnote{We have assumed that the proposed schemes use the same decoding method as RLC, e.g., Gaussian elimination.}, we can select a matrix from $Q^{(\epsilon)}$ using:
\begin{equation}
\label{eq.MS_LC}
\mathbf{P}=\arg\min_{\tilde{\mathbf{P}}\in Q^{(\epsilon)}}\sum_{\delta=0}^{q-1}\left(\mathrm{PoE}\!\left(\tilde{\mathbf{P}},\delta\right)-\frac{1}{q}\right)^2.
\end{equation}
Expression \eqref{eq.MS_LC}  ensures that each of the $q$ elements of $\mathbb{F}_q$ occupies a proportion of the elements of $\mathbf{P}$ that is as close as possible to $1/q$. We refer to this code design, which combines \eqref{eq.spark_maximization} and \eqref{eq.MS_LC}, as \textit{Maximum-Spark Linear Coding} (MS-LC). Given that $N=K+\epsilon$, the transmitter and the ground control station require $\epsilon$ realizations of the $\epsilon\times K$ matrix $\mathbf{P}$, that is, one realization of $\mathbf{P}$ for every possible value of $\epsilon$.

On the other hand, if we aim for a high-spark code that achieves the same decoding complexity as RLC, we can use the following expression to constrain membership in $Q^{(\epsilon)}$:
\begin{equation}
\label{eq.OS_PRLC}
\mathrm{PoE}\!\left(Q^{(\epsilon)},\delta\right)=\frac{1}{q}\quad\text{for all}\;\,\delta\in\mathbb{F}_q.
\end{equation}
In practice, we solve the optimization problem in \eqref{eq.spark_maximization} without any constraints but keep only a small number of matrices in $Q^{(\epsilon)}$. We then solve the optimization problem in \eqref{eq.spark_maximization} using \eqref{eq.OS_PRLC} as a constraint, and include the new solutions in $Q^{(\epsilon)}$. Each time $K$ source packets are input to the encoder, a matrix is selected uniformly at random from $Q^{(\epsilon)}$ and is assigned to $\mathbf{P}$. We refer to this coding scheme as \textit{Optimum-Spark Partly-Random Linear Coding} (OS-PRLC).

For example, let $Q^{(\epsilon)}$ contain all matrices from $\mathbb{F}_2^{\epsilon\times K}$ that maximize the spark of $\mathbf{H}^{\top}$.  Assume that \eqref{eq.MS_LC} identifies a matrix in $Q^{(\epsilon)}$ with a proportion of ones equal to $0.54$ for use in MS\nobreakdash-LC. The same matrix could also be used in OS\nobreakdash-PRLC. In that case, all other matrices would be removed from $Q^{(\epsilon)}$ but a second matrix should then be included in $Q^{(\epsilon)}$ with a proportion of ones equal to $0.46$. This constraint in the proportion of ones in the second matrix may prohibit matrix $\mathbf{H}^{\top}$ from achieving the maximum spark. Nevertheless, the proportion of ones in $Q^{(\epsilon)}$ will be $0.5$, thus \eqref{eq.OS_PRLC} will be satisfied. When $K$ source packets are input to the OS-PRLC encoder, one of the matrices in $Q^{(\epsilon)}$ will be chosen with equal probability and be used in the construction of $\mathbf{G}$ and $\mathbf{H}^{\top}$.

The following section compares RLC with MS-LC and OS-PRLC in terms of their achievable decoding probability, that is, the probability that all of the source packets will be recovered when only correctly received packets are considered, and when syndrome decoding is employed to repair corrupted packets.


\section{Results and Discussion}
\label{sec.Results}

We focus on a simple system in which $M=2$ drones collect and transport packets to a ground control station. The channels between the transmitter and the two drones are assumed to be independent and statistically similar, i.e., $\varepsilon_1=\varepsilon_2$. The probability that the ground control station will obtain a packet in error is given by $\varepsilon=\varepsilon_1\varepsilon_2$, which is the probability that both drones will receive the packet in error. In general, as the value of $M$ increases, the value of $\varepsilon=\prod_{m=1}^{M}\varepsilon_m$ reduces.

\begin{figure}[t]
\centering
\includegraphics[width=0.95\columnwidth]{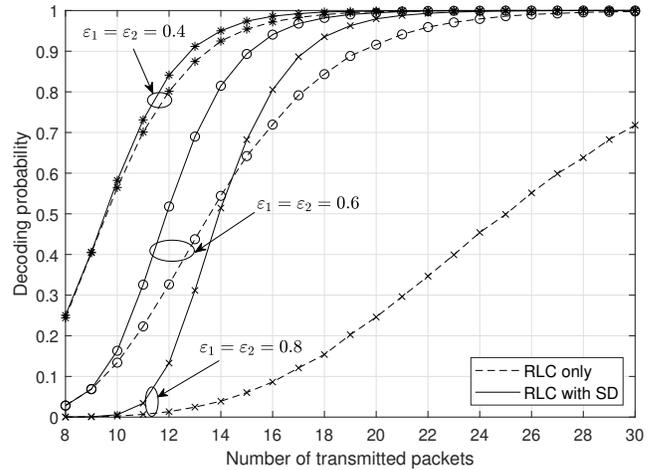}
\caption{Comparison between RLC and RLC with SD, in terms of the probability that the ground control station will recover $K=8$ source packets for an increasing number of transmitted packets $N$, when $M=2$ relay drones are used. The packet erasure probabilities take values from $\{0.4, 0.6, 0.8\}$.}
\label{fig:dec_prob_erasures}
\vspace{-2.6pt}
\end{figure}

As in \cite{Mohammadi2016}, we consider the encoding of $K=8$ source packets using systematic RLC over $\mathbb{F}_2$. The transmitter broadcasts the $N$ generated packets to the two drones over packet erasure channels with $\varepsilon_1=\varepsilon_2\in\{0.4,0.6,0.8\}$. The ground control station decodes the delivered packets using either an RLC decoder or a PPR-assisted RLC decoder. In the latter case, PPR employs syndrome decoding (SD), which was discussed in Section \ref{sec.Syndrome_Decoding}. As shown in Fig.~\ref{fig:dec_prob_erasures}, the performance gain of RLC with SD over RLC grows as the packet erasure probability increases. For example, when $\varepsilon_1=\varepsilon_2=0.8$ and $N=29$, the ground control station will recover the source packets with probability $68\%$, if the RLC decoder discards corrupted packets. However, if SD is used to repair corrupted packets, the same decoding probability will be achieved for only $N=15$ packet transmissions. If the value of $N$ is increased to $26$, the decoding probability of RLC with SD will rise to $99.9\%$.

As explained in Section~\ref{sec.Design}, matrix $\mathbf{P}$ in $\mathbf{H}^{\top}=[\,-\mathbf{P}\,|\,\mathbf{I}_{N-K}\,]$ is randomly generated when PPR-assisted RLC is used, thus maximization of the spark of $\mathbf{H}^{\top}$ is not guaranteed and the potential of PPR is not fully exploited. MS-LC uses \eqref{eq.spark_maximization} and \eqref{eq.MS_LC} to obtain deterministic designs of $\mathbf{P}$ that maximize the spark of $\mathbf{H}^{\top}$, while trying to keep a balance between the number of zeros and the number of ones in $\mathbf{P}$. The outcome of the search for maximum-spark realizations of $\mathbf{H}^{\top}$, when $K=8$ and $N$ varies from $9$ to $18$, is summarized in Table~\ref{tb:codes}. For each value of $N$, the maximum-spark matrix $\mathbf{H}^{\top}$ identified by the search process contains a submatrix $\mathbf{P}$ with a proportion of entries set to $1$ that is as close to half as possible. Contrary to MS-LC, OS-PRLC uses \eqref{eq.spark_maximization} and \eqref{eq.OS_PRLC} to identify not a single matrix $\mathbf{H}^{\top}$ but a set of matrices, for each value of $N$. This approach ensures that the two elements of $\mathbb{F}_2$ appear with the same frequency, on average, in $\mathbf{P}$. However, as is evident from Table~\ref{tb:codes}, not all of the matrices in each set always achieve the maximum spark. For example, when $N\in\{9,10,11,15,18\}$, the matrices in the sets used by OS-PRLC are all maximum-spark matrices; for the remaining values of $N$, constraint \eqref{eq.OS_PRLC} necessitates the inclusion of matrices in the sets that have spark lower than the maximum value.

\begin{table}[t]
\renewcommand{\arraystretch}{1.5}
\caption{Spark of matrix $\mathbf{H}^{\top}=[\,-\mathbf{P}\,|\,\mathbf{I}_{N-K}\,]$ and proportion of ones in matrix $\mathbf{P}$ for $q=2$, $K=8$ and $N=9,\ldots,18$.}%
\centering
\begin{tabular}{|c||c|c||c|c|}%
\hline
\parbox[t]{2mm}{\multirow{2}{*}{\textit{N}}}
& \multicolumn{2}{c||}{MS-LC} & \multicolumn{2}{c|}{OS-PRLC} \\  \cline{2-5}
& Prop. of $1$ & Spark & Lowest spark & Highest spark \\
\hline\hline
 $9$  & $0.5$    & $1$ & $1$ & $1$ \\
$10$  & $0.5$    & $2$ & $2$ & $2$ \\
$11$  & $0.5$    & $2$ & $2$ & $2$ \\
$12$  & $0.625$  & $3$ & $2$ & $3$ \\
$13$  & $0.6$    & $4$ & $3$ & $4$ \\
$14$  & $0.5417$ & $4$ & $3$ & $4$ \\
$15$  & $0.5179$ & $4$ & $4$ & $4$ \\
$16$  & $0.5781$ & $5$ & $4$ & $5$ \\
$17$  & $0.5139$ & $5$ & $4$ & $5$ \\
$18$  & $0.5$    & $5$ & $5$ & $5$ \\
\hline
\end{tabular}
\label{tb:codes}
\vspace{-2.6pt}
\end{table}

\begin{figure}[t]
\centering
\includegraphics[width=0.92\columnwidth]{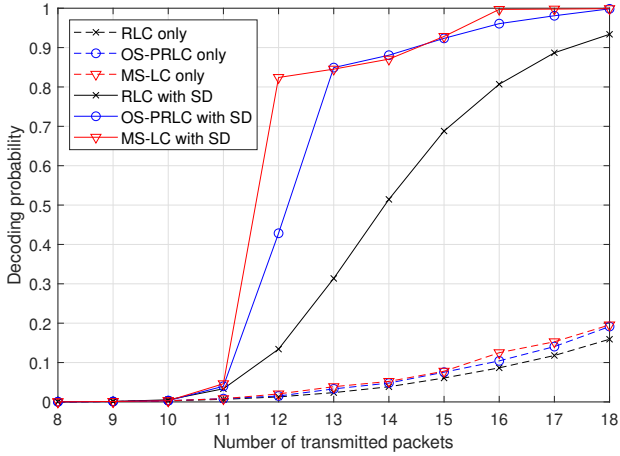}
\caption{Comparison of the decoding probabilities of RLC, MS-LC and OS-PRLC for $M=2$ drones, packet erasure probabilities $\varepsilon_1=\varepsilon_2=0.8$, $K=8$ source packets and an increasing number of transmitted packets $N$.}
\label{fig:dec_prob_SD}
\vspace{-3pt}
\end{figure}

The decoding probabilities of RLC and RLC with SD for $\varepsilon_1=\varepsilon_2=0.8$, which have been plotted in Fig. \ref{fig:dec_prob_erasures}, have also been included in Fig. \ref{fig:dec_prob_SD} and are used as benchmarks for the decoding probabilities of MS-LC and OS-PRLC. We observe that both MS-LC and OS-PRLC achieve marginally higher decoding probabilities than RLC, when PPR is not used. The random selection and combination of source packets in RLC does not exclude the possibility that a source packet is never selected in the encoding process. This event is translated into an all-zero column in $\mathbf{P}$ and has a negative impact on the decoding probability of RLC. For $N>K+1$, maximization of the spark of $\mathbf{H}^{\top}$ prevents the occurrence of all-zero columns\footnote{The spark of $\mathbf{H}^{\top}$ would be $1$ if $\mathbf{P}$ had one or more all-zero columns.} in $\mathbf{P}$ and provides a small advantage to MS-LC and OS-PRLC over RLC. The impact of spark optimization on the decoding probability can be better appreciated when PPR, in the form of SD, is activated. As illustrated in Fig.~\ref{fig:dec_prob_SD}, the decoding probabilities of MS-LC and OS-PRLC increase more sharply than that of RLC, when all three schemes employ~SD. MS-LC performs better than OS-PRLC because it prioritizes decoding success over decoding complexity, as explained in Section~\ref{sec.Design}.

\begin{figure}[t]
\centering
\includegraphics[width=0.92\columnwidth]{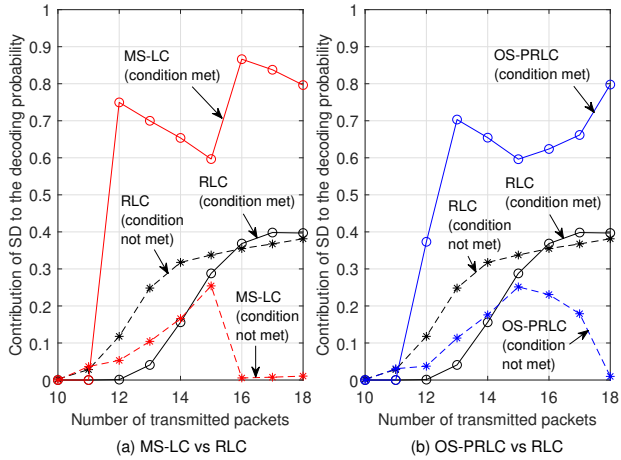}
\caption{Contribution of SD to the decoding probability of RLC, MS-LC and OS-PRLC when the condition in Theorem~\ref{th.uniqueness} is met and when it is not met.}
\label{fig:SD_contributions}
\vspace{-3pt}
\end{figure}

For a given number of transmitted packets in Fig.~\ref{fig:dec_prob_SD}, the difference between the value of the decoding probability of a scheme that uses SD (e.g., RLC with SD) and the value of the decoding probability of the same scheme when SD is not used (e.g., RLC only) provides the probability of SD successfully assisting that scheme where the stand-alone scheme would have failed. This contribution of SD to the overall decoding probability of RLC, MS-LC and OS-PRLC has been decomposed into two probabilities in  Fig.~\ref{fig:SD_contributions}: (i) the probability of SD being successful because the condition in Theorem~\ref{th.uniqueness} has been met and the unique sparse solution has been determined, and (ii) the probability of SD being successful because the identified solution, although not unique, estimated some of the rows of the error matrix correctly and increased the number of recovered source packets to $K$. Fig.~\ref{fig:SD_contributions} shows that the former probability increases significantly when the condition in Theorem~\ref{th.uniqueness} is met, not by chance as in RLC but by design as in MS-LC and OS-PRLC.

The sawtooth shape of the MS-LC curve in Fig.~\ref{fig:SD_contributions}(a) is of interest. Table~\ref{tb:codes} shows that when $N$ rises from $11$ to $12$, the spark of $\mathbf{H}^{\top}$ increases from $2$ to $3$. Fig.~\ref{fig:SD_contributions}(a) confirms that this increase in the spark leads to a higher decoding probability because the chances of the condition in Theorem~\ref{th.uniqueness} being met have improved. As $N$ grows from $12$ to $15$, the nonzero elements in the error matrix increase at a rate that reduces the chances of increments in the spark of $\mathbf{H}^{\top}$ maintaining a high decoding probability. Only when the spark of $\mathbf{H}^{\top}$ reaches $5$ for $N=16$, the condition is satisfied more frequently and the decoding probability improves.


\section{Conclusions}
\label{sec.Conclusions}

A delay-tolerant network composed of a remote transmitter broadcasting information to drones, which store it and deliver it to a ground control station, is expected to offer high data reliability while operating in challenging wireless conditions. The ground control station will decode the information delivered by the drones with a high probability if random linear coding (RLC) is combined with partial packet recovery (PPR). We explained that the effectiveness of PPR depends on the spark of the transpose of the parity-check matrix of the code. We argued that PPR-assisted RLC does not fully exploit the potential of PPR because of the random design of the parity-check matrix of RLC. Based on the assumption that feedback channels are unavailable, we presented two deterministic designs of parity-check matrices. We demonstrated through simulations that the proposed schemes achieve a comparable or better performance than RLC, when decoding is not assisted by PPR. When PPR is used, both schemes can decode the delivered information with a noticeably higher probability than RLC.


\bibliographystyle{IEEEtran}
\bibliography{IEEEabrv,references}


\end{document}